\documentclass{aa}
\usepackage{color,graphics}
\begin{document}
\definecolor{light}{gray}{0.9}

   \thesaurus{02     
              (11.08.1;  
               12.03.3;  
               12.04.1;  
               12.07.1)} 
   \title{Microlensing results from APO 
          monitoring of the double quasar Q0957+561A,B between 1995 and 1998}

   \titlerunning{Microlensing results from 4-years of APO 
                 data on quasar Q0957+561A, B}

   \author{	J.~Wambsganss\inst{1}\fnmsep\inst{3}\fnmsep\inst{4},
		R.~W.~Schmidt\inst{1}\fnmsep\inst{2},
           	W.~Colley\inst{5}, 
		T.~Kundi\'c\inst{6} \and
           	E.~L.~Turner\inst{7}}

   \authorrunning{Wambsganss et al.}

    \offprints{J.~Wambsganss}

   \institute{
		Universit\"at Potsdam,
              Institut f\"ur Physik,
              Am Neuen Palais 10,
              14469 Potsdam,
              Germany,
		{\tt jkw@astro.physik.uni-potsdam.de}
              \and              Institute of Astronomy,
              University of Cambridge,
              Madingley Road,
              Cambridge CB3 0HA,
              UK,
		{\tt 	rschmidt@ast.cam.ac.uk}
              \and
		Max-Planck-Institut f\"ur Gravitationsphysik,
		Am M\"uhlenberg 1,
		14476 Golm,
		Germany
              \and
		The University of Melbourne,
		School of Physics,
		Parkville, Vic 3052, 
		Australia
              \and
              Harvard-Smithsonian Center for Astrophysics,
              60 Garden Street,
              Cambridge MA 02138
              USA,
		{\tt wcolley@barbecue.harvard.edu}
              \and
              Renaissance Technologies,
              600 Route 25A,
              East Setauket,
              NY 11733,
              USA,
		{\tt tomislav@rentec.com}
              \and
              Princeton University Observatory,
              Peyton Hall,
              Princeton,
              NJ 08544,
              USA,
		{\tt elt@astro.princeton.edu}
             }

   \date{Received \today; accepted }

   \maketitle

   \begin{abstract}
If the halo of the lensing galaxy 0957+561 is made of massive
compact objects (MACHOs),
they must affect the lightcurves of the quasar images Q0957+561 A and B
differently.
We search for this microlensing effect in the double quasar by
comparing  monitoring data for the two images A and B -- 
obtained with the 3.5m Apache Point Observatory from 1995 to 1998  --
%
%
with intensive
numerical simulations. This way we test whether the
halo of the lensing galaxy can be made of MACHOs of various masses.
We can exclude a halo entirely made out of MACHOs with  masses 
between $10^{-6}\,M_{\odot}$ and $10^{-2}\,M_{\odot}$  for quasar sizes of 
less than $3\times 10^{14}\,h_{60}^{-1/2}$\,cm,
hereby extending previous limits upwards by one order of magnitude.
      \keywords{gravitational lensing --
	dark matter  --
	quasars: individual: Q0957+561 --
	galaxies: halos -- 
	cosmology: observations 
	}
   \end{abstract}

\section{Introduction}
The nature of the dark matter is still one of the most pressing questions
in current astrophysics. 
Many experiments are running or being 
built/planned in order to test the existence of various particle physics 
candidates,
like wimps, axions, or neutralinos (for a review on the particle physics
aspects of dark matter, see Raffelt 1997).
There are also intense searches underway for astrophysical dark matter
candidates, like black holes (from stellar to galactic mass scales),
brown dwarfs, or other compact objects.

Gott (1981) and Paczy\'nski (1986) put forward the idea
to use gravitational lensing
for the search of compact dark matter objects in galactic halos. Gott (1981)
suggested to look for fluctuations in the lightcurves of multiply imaged
quasars in order to detect compact objects along the line of 
sight in the halo of the lensing galaxy, 
which had been shown earlier by Chang and Refsdal (1979) to have observable
consequences.
Paczy\'nski (1986) proposed to  probe the content 
of compact dark objects in the Milky Way halo 
by monitoring the brightnesses of about 10$^7$  individual
stars in the Large Magellanic Cloud (LMC). 
The  latter paper led to a
whole industry of observing projects, like MACHO (Alcock et al. 2000) and
EROS (Lasserre et al. 2000).
Gott (1981)'s suggestion to study the halos of lensing galaxies by way of 
monitoring background quasars -- though in principle as powerful 
a method -- was pursued, however, with much less fervor 
in terms of people involved,
total observing nights or CPU-time used.

We followed the latter method and
present here results on the possible matter contents of the halo of
the lensing galaxy in the lens system 0957+561. 
They are based on comparing 
lightcurves of the gravitationally lensed 
double quasar Q0957+561A,B, 
obtained by four years of monitoring (1995 - 1998) at
the 3.5m-Apache Point Observatory 
(Colley, Kundi\'c \& Turner 2000, hereafter CKT) 
with extensive numerical simulations.

\section{Data and Method}

\subsection{The double quasar Q0957+561 and the data}
The double quasar Q0957+561A,B was the first multiply imaged quasar
discovered (Walsh, Carswell \& Weymann 1979). 
It consists of two quasar images of $R \approx 16$\,mag and identical
redshift $z=1.41$, separated by $\Delta \theta = 6.1''$. 
Image A is about 5 arcseconds away from the center of the lensing galaxy,
image B is about 1 arcsecond off.
Very briefly after its discovery, Chang \& Refsdal
(1979) suggested that stellar mass objects in the light path of one of the
images can produce uncorrelated changes in its apparent brightness.
The lensing galaxy at $z_{\rm G} = 0.36$ is the central galaxy of a 
rich cluster,   whose
weak lensing effects on background galaxies
have been seen (Barkana et al. 1999).
Q0957+561A,B  is the best investigated gravitational
lens with far more than 100 papers written about it. 
The time delay between the two images has been established to be
around 417 days (e.g., Schild \& Thomson 1997; Kundi\'c et al. 1998).
Earlier results on microlensing had shown that for a period of
about 5 years, there was a monotonic change between the
time-delay-corrected
apparent brightnesses of images A and B (Schild 1996, Pelt et al. 1998,
Refsdal et al. 2000, Gil-Merino et al. 1998). 
The time and amplitude
of that fluctuation is consistent with microlensing due
to low mass stars,  which is expected to happen again for image B 
before too long.

\begin{figure}[htb]
	\resizebox{\columnwidth}{!}{\includegraphics{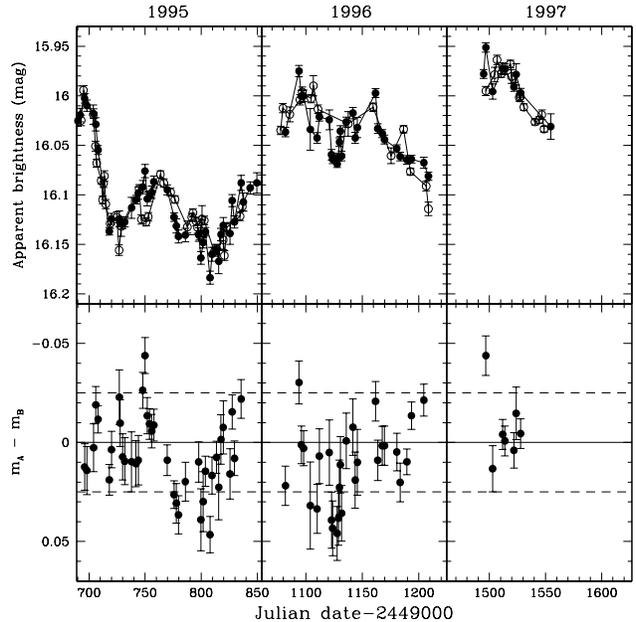}}
	\caption{Top: Lightcurves of image A (solid) and image B (open, 
	shifted in time by $\Delta t = -417$ days  
	and in magnitude by $\Delta m_{\rm AB} = 0.125/0.158$ mag). Bottom: 
	``Difference lightcurve'' between quasar images A and 
	time- and magnitude shifted B.
	In order to guide the eye, dashed lines are
	drawn at differences of +0.025\,mag and $-$0.025\,mag. Julian dates
	correspond to image A data points.  }
	\label{Fig_data}
	\end{figure}

In Fig.~\ref{Fig_data} we show the lightcurves of images A and B
covering the epochs 1995-1997 for the leading image A (and epochs
1996-1998 for the trailing image B), 
based on data by  CKT.
The 1995/1996 data set 
has already been analysed with respect to microlensing
by Schmidt \& Wambsganss (1998, hereafter SW98).
Analysing the complete data set which covers three full years
instead of 160 days allows us here
to extend the mass limits by one order of magnitude.
The difference light curve in the lower panel of Fig.~\ref{Fig_data}
has been calculated by interpolating the data points of image B and
subtracting them from the corresponding data point of image A. We
have only included data points in the difference lightcurve where an
interpolation was possible.
It is obvious from the lower part of Fig.~\ref{Fig_data} that there
are no major 
effects in the difference lightcurve.
There are some trends visible, but it is not clear whether they can
be attributed to microlensing or whether they are due to some other
systematic effects.
The dip around day 1125, for example, is likely to be the effect of a
lack of data points for quasar B to interpolate in between.
In any case, including the 1-$\sigma$ error
bars, all the data points are consistent with the conservative 
assumption that no microlensing with an amplitude 
$|\Delta m| > 0.05$\,mag has been detected.

%
%
%
%
%
%
%
%
	\begin{figure}[htb]
	\begin{minipage}{\columnwidth}
	\resizebox{4.2cm}{!}{\includegraphics{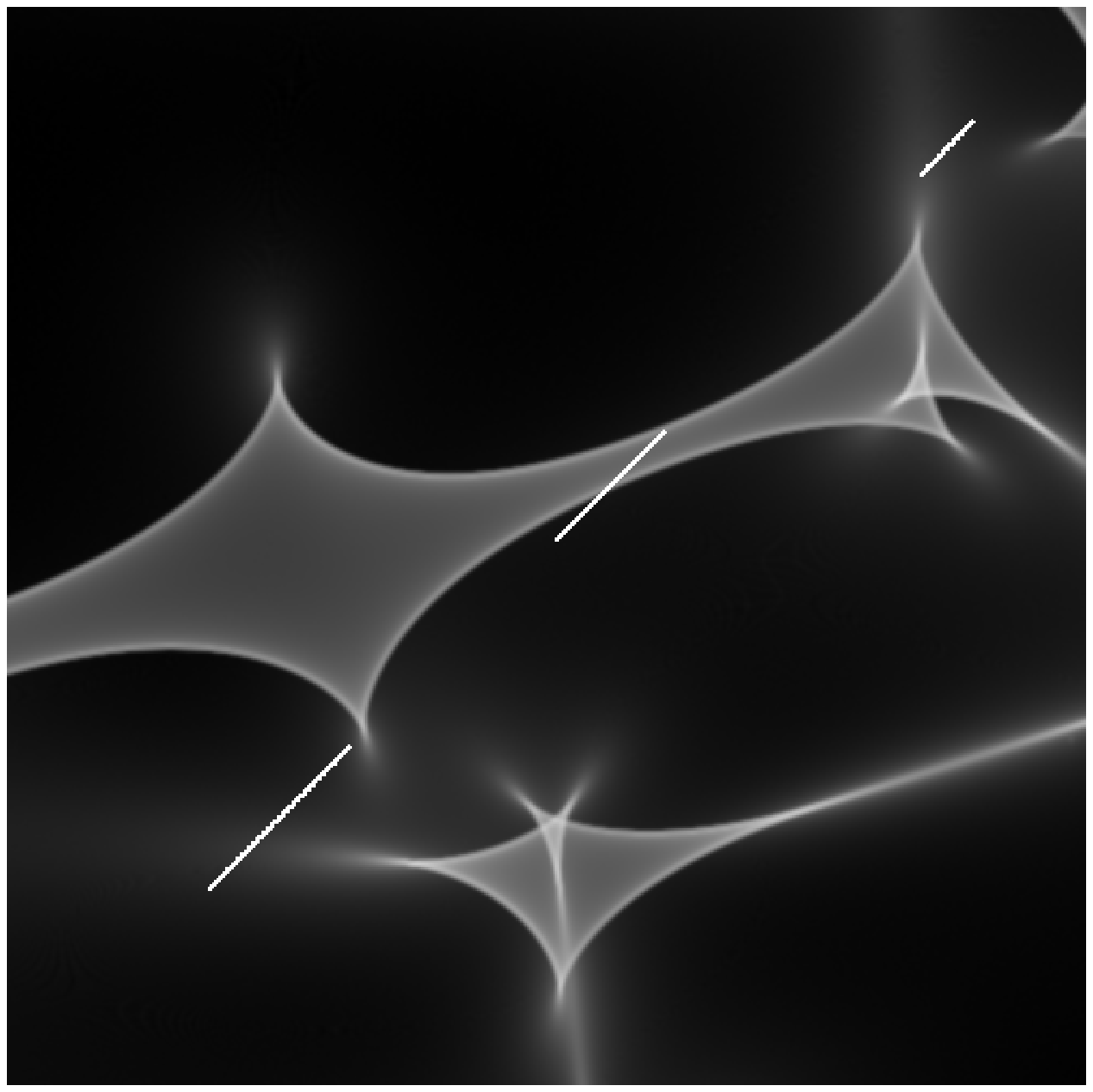}}
	\hfill
	\resizebox{4.2cm}{!}{\includegraphics{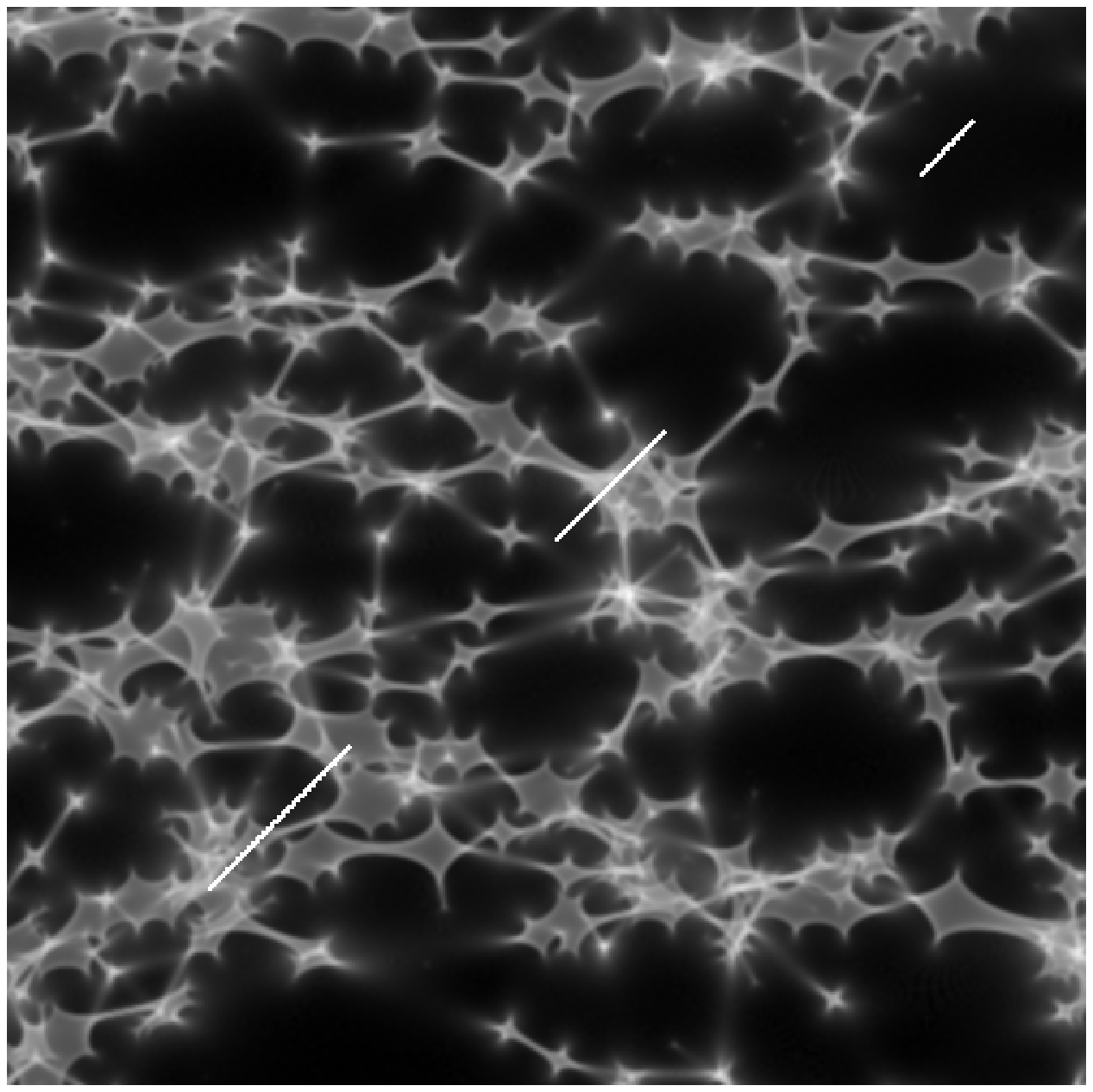}}
	\end{minipage}
	\vspace{0.2cm}
%
%
%
%
\begin{center}
	\resizebox{7.8cm}{!}{\includegraphics{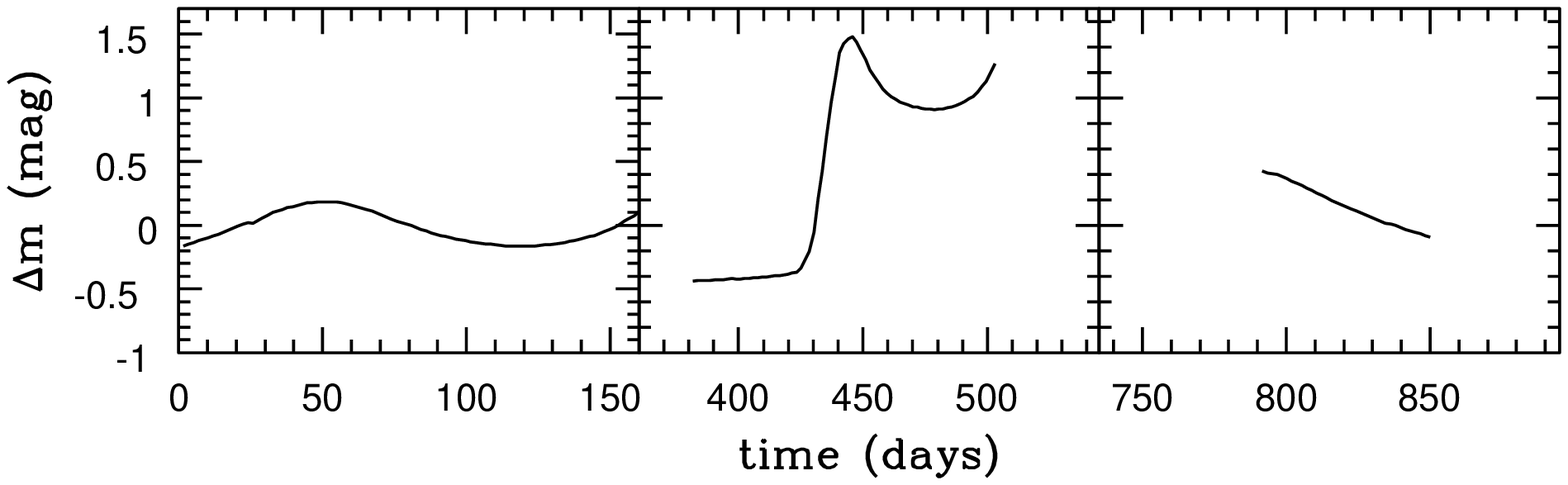}}

	\resizebox{7.8cm}{!}{\includegraphics{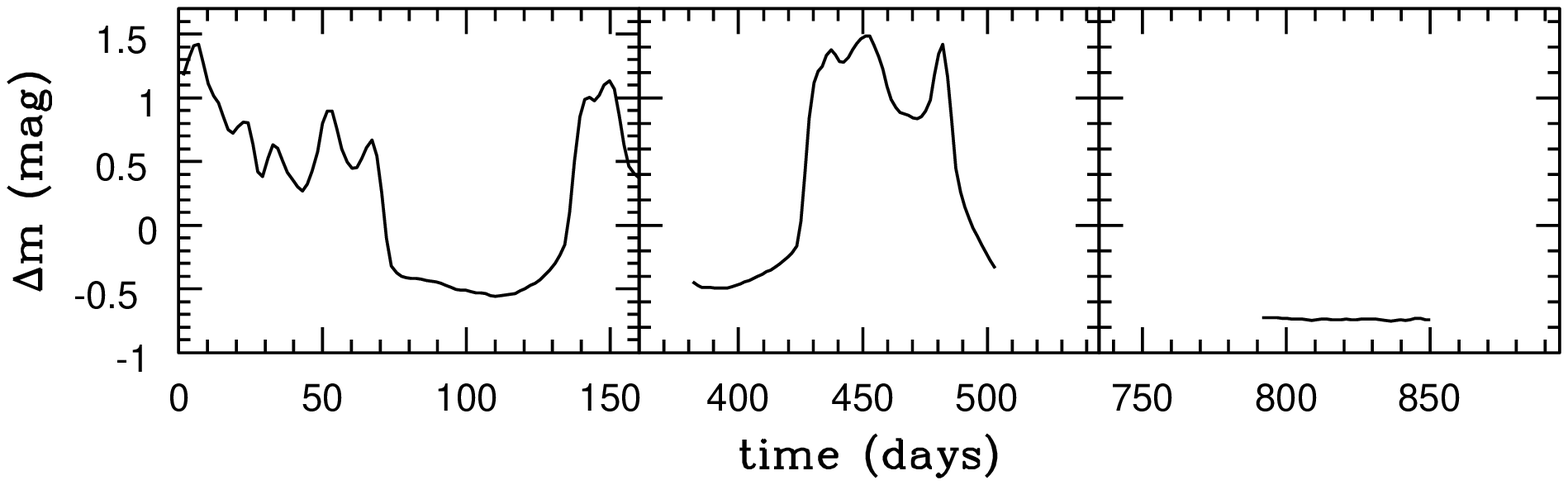}}
\end{center}

	\caption{Top: Small parts of microlensing magnification patterns 
	corresponding to $m_{\rm macho}=10^{-3}M_\odot$ (left) 
	and $10^{-5}M_\odot$ 
	(right), sidelengths are 4 $r_{\rm E}$.
	The three-part straight lines corresponds to a random track 
	with sampling modelled after the actual data (cf. 
	Fig.~1).
	Bottom:
	Microlensing lightcurves corresponding to the above tracks 
	(quasar size $\sigma_{\rm Q} = 3 \times 10^{13}$cm; 
	effective transverse velocity $v_\perp = 600$km/sec).
	}
	\label{Fig_light}
	\end{figure}

\subsection{Numerics}
The method we use is described in detail in 
SW98 and in Schmidt (2000).
The first step is to produce a ``difference lightcurve'' from  the 
data of images  A and B, by shifting one set by
the appropriate time delay (we chose $\Delta t =  417$ days, 
see Kundi\'c et al. 1998) 
and by the difference in apparent brightness
($\Delta m_{\rm AB} = 0.125/0.158$ mag\footnote{$\Delta m_{\rm AB}$
	changed due to a change in the mirror's reflectivity 
	after realuminization, see CKT}).
The second step is to either identify  significant
fluctuations in the difference lightcurve which could be attributed
to microlensing, or to find 
an upper limit on the possible action of microlenses.

In the third step we produce magnification patterns
via numerical simulations for both images
A and B with the appropriate values of surface mass density $\kappa$
and external shear $\gamma$ (cf. Schneider, Falco \& Ehlers 1992).
We investigated three different scenarios, 
assuming 100\%, 50\% or 25\%  of the matter density consists of Machos,
respectively.
We explore mass ranges from $10^{-7} \le m_{\rm macho}/M_\odot \le 1.0$.
In the fourth step,
the resulting magnification patterns are
convolved
with a brightness profil of the quasar.
We chose Gaussian profiles with sizes of 
$\sigma_{\rm Q}/cm = 10^{14}, 3\times 10^{14}, 10^{15}, 3\times 10^{15}$
(the smallest size we consider corresponds to a few Schwarzschild radii
of a presumed central supermassive 
black hole with a mass of about $10^8$ M$_\odot$).
Step 5  is the simulation of  randomly oriented (straight) tracks 
through these magnification patterns that cover the same sampling
intervals as the actual observations. 
In the final step 6 we determine the fraction of all lightcurves
for a particular parameter pair $m_{\rm macho}, \sigma_{\rm Q}$ 
that produced fluctuations
larger than the ones observed.

For the ray shooting simulations (cf. Wambsganss 1999),
we used values of 
$\kappa_{\rm A} = 0.32, \gamma_{\rm A} = 0.18$ and
$\kappa_{\rm B} = 1.17, \gamma_{\rm B} = 0.83$ for convergence and 
shear of images A and B, as in SW98.
For each of the masses 
$m_{\rm macho}/M_\odot = 10^{-7}$, 
            $10^{-6}$,  ..., 
            $10^{-1}$, 
            $1$, we used three independent magnification
patterns with 2048$^2$ pixels each, and sidelengths of L = 16, 160, 1600 $r_{\rm E}$
(where the Einstein radius in the source plane for a 1 $M_\odot$-object is 
$r_{\rm E} = \sqrt{   { {4 G m} \over c^2} { {D_{\rm s} D_{\rm ds}} \over D_{\rm d}}} 
\approx 4.8 \times 10^{16} \sqrt{m/M_{\odot}} h_{60}^{-0.5}$ cm; 
$D_{\rm d}$, $D_{\rm s}$, $D_{\rm ds}$ are the angular diameter distances
observer-lens, observer-source, lens-source, respectively, 
$c$ is the velocity of light, and $G$ is the gravitational constant).
We assumed an effective transverse velocity of $v_\perp = 600$ km/sec
		(cf. Paczy\'nski 1986; Kayser, Refsdal, Stabell 1986).
For each parameter pair $m_{\rm macho}$,  $\sigma_{\rm Q}$, 
we produced 100,000 simulated microlensing lightcurves 
each for image A and image B,
sampled like the observed data set. 
We looked for the differences between the lowest and highest part
in each lightcurve and binned those $10^5$ maximum differences.
The fractions of those lightcurves  that 
showed fluctuations larger than the observed difference of 
$\Delta m_{\rm obs} = 0.05$\,mag 
were labelled $p_{\rm A}$ and $p_{\rm B}$, respectively.
We assumed the fluctuations to be independent between images A and B,
i.e. the combined
exclusion probability is defined as\footnote{This is not exactly 
	identical to detecting a certain $\Delta m$ in the
	difference lightcurve (e.g., two large but opposite changes
	could just cancel); but for our purposes it is close enough.
}
$p_> = 1 - (1 - p_{\rm A})(1-p_{\rm B})$.
The method is illustrated in Fig.~\ref{Fig_light}, 
where two panels with small parts of the magnification
patterns are reproduced for objects of masses 
$m_{\rm macho} = 10^{-3}$ (left), and $10^{-5}$ (right).
The three white line segments show how the tracks were
modelled after the time coverage of the real data 
in Figure \ref{Fig_data}. 
The corresponding microlensed lightcurves are shown in the lower part
of Figure \ref{Fig_light}.

\section{Results,  Discussion and Summary}

\begin{figure}[htb]
	\resizebox{9.0cm}{!}{\includegraphics{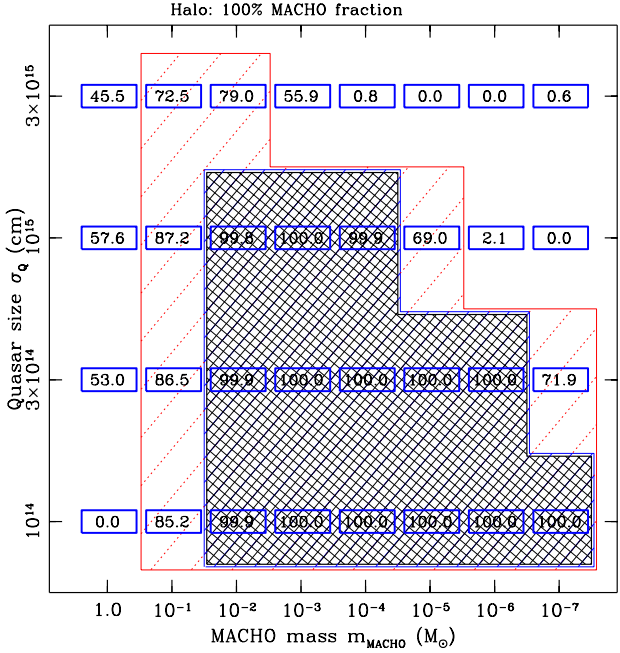}}
	\caption{Exclusion Probability (in percent) as a function of 
	Macho mass $m_{\rm macho}$ and 
	quasar size $\sigma_{\rm Q}$,
	for 100\% of the halo mass in MACHOs.
	The shaded parts encircle regions of parameter
	space that can be excluded at the
	67\%, 95\%, and 99.7\% probability level
	(increasing line density;
	the latter two coincide).
	}
	\label{Fig_lego_1a}
	\end{figure}

\begin{figure}[htb]
	\resizebox{8.7cm}{!}{\includegraphics{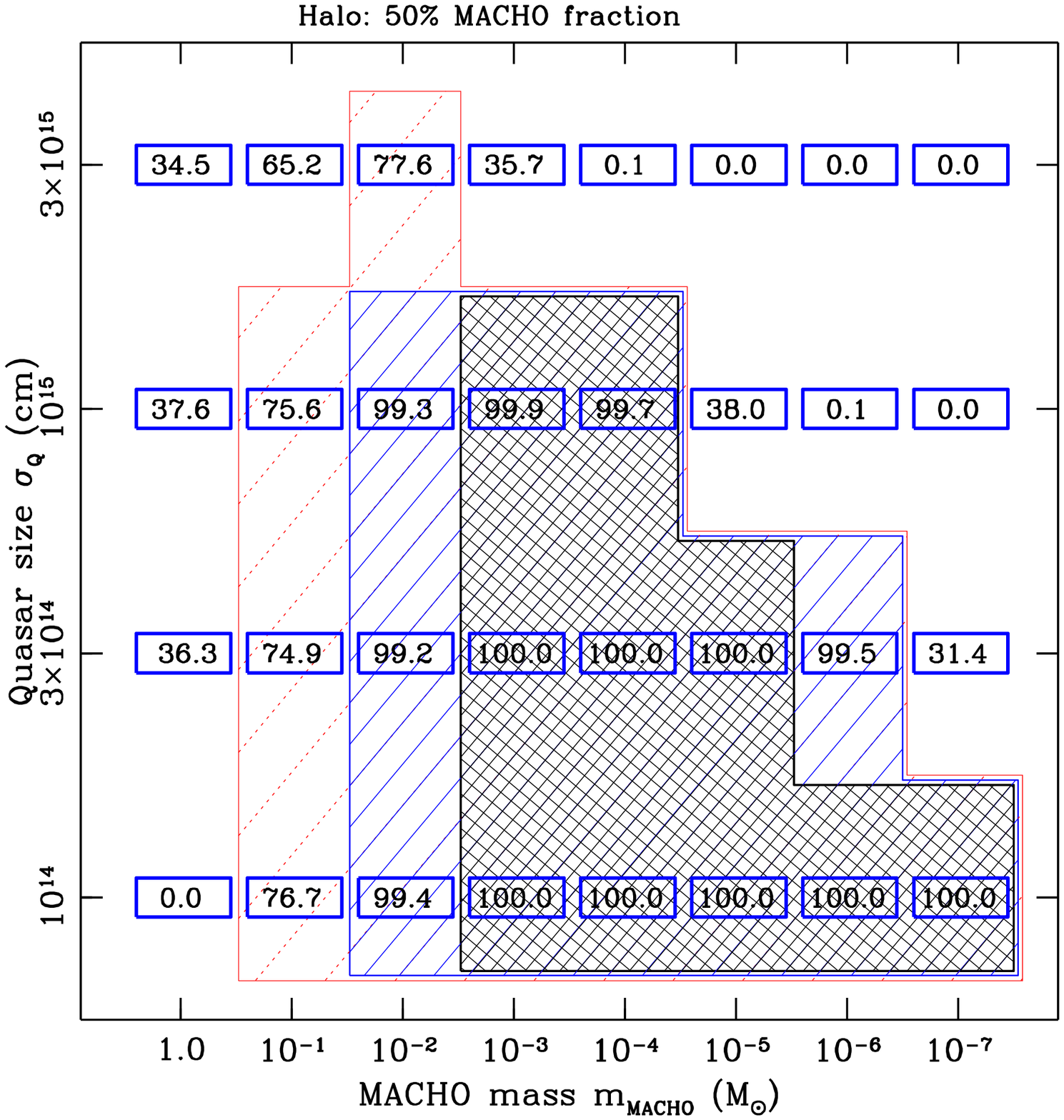}} 
	\caption{Same as Fig.~\ref{Fig_lego_1a} 
	for 50\% of the halo mass in MACHOs.}
	\label{Fig_lego_1b}
	\end{figure}

\begin{figure}[htb]
	\resizebox{8.7cm}{!}{\includegraphics{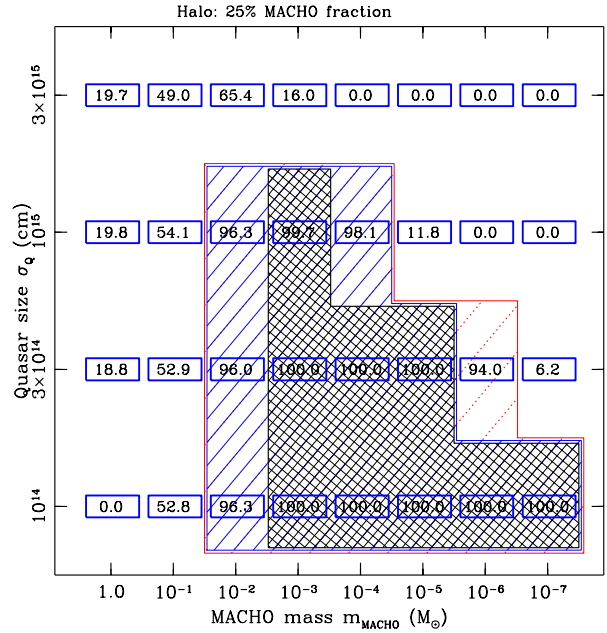}}
	\caption{Same as Fig.~\ref{Fig_lego_1a}
		for 25\% of the halo mass in MACHOs.}
	\label{Fig_lego_1c}
	\end{figure}

The lightcurves of Q0957+561A, B as monitored between 1995 and 1998
do not show any significant differences beyond $\Delta m = 0.05$\,mag,
when corrected for time delay and magnitude difference.
In Fig.~\ref{Fig_lego_1a} we present the resulting 
``exclusion probabilities'' for 
the two parameters ``Macho mass'' $m_{\rm macho}$ and 
quasar size $\sigma_{\rm Q}$ (assuming 100\% of the halo mass
is in MACHOs), 
derived from four years of monitoring
and comparison with numerical simulations.
The numbers indicate the percentage of the 100,000 simulated
lightcurves that showed fluctuations larger than the observed ones.
In the diagram, we encircled and shaded 
the parts of parameter space that produced exclusion probabilities
of 67\% (1-$\sigma$, thin shading),
   95\% (2-$\sigma$, medium shading), and
   99.7\% (3-$\sigma$, cross shading).
It is obvious that  a mass range from 
$10^{-6} \le m_{\rm macho}/M_\odot \le 10^{-2}$ can be 
excluded at the 3-$\sigma$ level for all 
quasar sizes except for the largest one, for which
$10^{-2}\,M_\odot$ is barely allowed. 
For objects of mass 0.1 $M_\odot$, the exclusion
probability is at the 70\% to 85\% level. We need a longer time
coverage, in order to improve significantly on these limits. 
In Figs.~\ref{Fig_lego_1b} and \ref{Fig_lego_1c} 
we present the resulting 
numbers and respective exclusion regions  for the assumption of
only 50\% or 25\% of the halo mass in MACHOs.
The exclusion probabilities get slightly smaller in these
cases, and the exclusion regions shrink a bit as well, but the
general picture does not change much.

%
%
%
%
%
%
%
%
%
%
%
%

Pelt et al. (1998)  and Refsdal et al. (2000) investigated the double 
quasar lightcurve as well. Their main focus was 
microlensing on
medium and long time scales (baseline 15 years), including 
a continuous change in the difference lightcurve of about 0.25 mag in 
the first five years.
Whereas their data set
covers a longer baseline than ours, we have many more and more accurate
data on short time scales ($\le$ 100 days). 
Furthermore, we also take into account 
the       gaps in the difference lightcurve, which  means
we treat the short term behaviour more realistically.
Pelt el al. (1998)'s  finding  that the quasar size is 
about $3\times 10^{15}$cm is consistent with 
our results (though we cannot put an upper limit on the size). 

We extend the ``exclusion'' area by roughly one order of
magnitude in mass, compared to the first results in SW98 and 
Wambsganss \& Schmidt (1998). 
This is due to the fact that the coverage of the
difference light curve increased by more than a factor of 4 
(without showing any more variability), 
and the mass limits increase with the square of 
the length scale. 
But it also means that in order to increase the 
limits from short/medium term microlensing by another order 
of magnitude in mass -- reaching the very interesting regime of solar 
mass objects -- the frequent monitoring has to continue for
another six or eight years.

The  method and results described here, in particular the 
exclusion diagram 
(Fig.~3; see also Fig.~2 of Refsdal et al. 2000), are very similar
to those of the groups investigating microlensing of the 
Milky Way halo 
(e.g., Alcock et al. 2000, Lasserre et al. 2000). 
Hence it is obvious that monitoring multiple quasars  
(Gott 1981) is as powerful a tool
in constraining the abundance of MACHOs in 
galactic halos as is monitoring LMC stars
(Paczy\'nski 1986).

\begin{acknowledgements}
   This work was supported in part by the Deut\-sche
   For\-schungs\-ge\-mein\-schaft (DFG), project number WA~1047/2-1,
   by the National Science Foundation (NSF), grant AST98-02802,
   and by the Australian Research Council (ARC, grant X00001713).
	It is a pleasure to thank the APO observers who helped
	obtaining this data, in particular 
	K. Gloria, N. C. Hastings,
	D. Long, E. Bergeron, C. Corson, and R. McMillan.
	We also like to thank Rolf Stabell for his careful reading and his
	comments which helped to improve the paper.
\end{acknowledgements}

\end{document}